\documentclass[aps,prb,twocolumn,groupedaddress]{revtex4}
\usepackage{epsfig}
\usepackage[american]{babel}
\usepackage{amsmath}
\usepackage[dvips]{color}
\usepackage{graphics}

\usepackage[latin2]{inputenc}
\usepackage[T1]{fontenc}
\usepackage[american]{babel}
\usepackage{color}
\usepackage{graphics}

\newcommand{\GRe}{\mbox{Re}}

\begin{document}

\title{Comparative study of organic metals and high-T$_c$ cuprates}

\author{S. Bari\v si\' c}

\affiliation{Department of Physics, Faculty of Science, University of
Zagreb, Bijeni\v cka c. 32, HR-10000 Zagreb, Croatia}

\author{O. S. Bari\v si\' c}

\affiliation{ Jo\v zef Stefan Institute, SI-1000 Ljubljana, Slovenia and\\Institute of Physics, Bijeni\v cka c. 46, HR-10000 Zagreb, Croatia}

\begin{abstract}

The Bechgaard salts and the high T$_c$ cuprates are described by two and three
band models, respectively, with the lowest band (nearly) half filled. In
organics the interactions are small, while in cuprates the repulsion $U_d$ on
the Cu site is the largest energy. The Mott AF state is stable in undoped
materials in both cases. In the metallic phase of cuprates the
$U_d\rightarrow\infty$ limit produces a moderate effective repulsion. The
theories of the coherent SDW and charge-transfer correlations in the metallic
phases of organics and cuprates are thus similar. In (undoped) organics those
correlations are associated with commensurate $2k_F$ SDW and $4k_F$ bond or
site modes. The corresponding modes in metallic cuprates are the
incommensurate SDW and in particular O$_x$/O$_y$ quadrupolar charge transfer
with wave vector $2q_{SDW}=q_0+G$. They are enhanced for dopings $x>0$, which
bring the Fermi level close to the van Hove singularity. Strong coupling to
the lattice associates the static incommensurate O$_x$/O$_y$ charge transfer
with collinear ``nematic'' stripes. In contrast to organics, the coherent
correlations in cuprates compete with local $d_{10}\leftrightarrow d_9$
quantum charge-transfer disorder.

\end{abstract}

\maketitle

\section{Introduction}

The high-T$_c$ cuprates are usually described by the Emery model \cite{em1}
where the role similar to the external dimerization \cite{bs7} in the
Bechgaard salts is played by the Cu-O hopping $t_{pd}$, which puts two oxygens
in the CuO$_2$ unit cell and makes the lowest of three bands half filled. The
weak coupling theory at zero doping $x$ is then a quite straightforward
analogue \cite{fr2,fr3,dz2,bs5} of the 1D theory, provided that the imperfect
nesting associated with the O$_x$-O$_y$ hopping $t_{pp}$ is ignored. The
appearance of the $x=0$ Mott-AF state is essentially independent of the value
of $\Delta_{pd}= \varepsilon_p-\varepsilon_d>0$ where $\varepsilon_d$ and
$\varepsilon_p$ are the Cu and O site energies in the hole picture,
respectively.

\begin{figure}[tb]

\begin{center}{\scalebox{0.35}
{\includegraphics{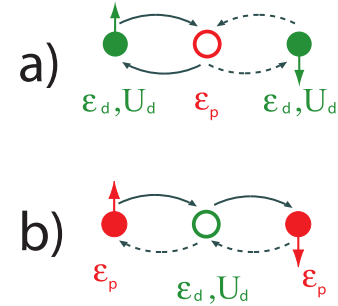}}}
\end{center}

\caption{(Color online) (a) Superexchange of two holes on Cu sites via the
empty oxygen site; (b) scattering of two holes on the O-sites via the empty
Cu-site.\label{Fig1}}

\end{figure}

However, while the Hubbard repulsion in the Bechgaard salts is small, in the
high-T$_c$ cuprates the repulsion $U_d$ on the copper site is the largest
energy \cite{em1}. It has long been maintained \cite{fr2,fr3} that the
fundamental question in the high-T$_c$ cuprates concerns the nature of
correlations which reduce $U_d$. The relevant $U_d=\infty$ limit is usually taken starting from the unperturbed state with average Cu-occupation $n_d^{(0)}=1$. The lowest order process shown in Fig.~\ref{Fig1}a then corresponds to the fact that two holes on the neighboring Cu-sites can hop simultaneously to the intermediate O-site, empty at $x=0$, provided that their spins are opposite. This leads to the superexchange \cite{za1} $J_{pd}\sim t_{pd}^4/\Delta_{pd}^3$ which is the basis of the $t-J$ models \cite{za1}. Alternatively, one can assume that all holes in the unperturbed metallic state are on the O-sites, i.e., that $n_d^{(0)}=0$. When two $p$-holes of opposite spin are crossing the intermediate empty Cu-site one hole hops to the Cu-site when $t_{pd}$ is turned on, while the other has to wait as long as the Cu-site is occupied. The waiting time is of the order of  $(\varepsilon_d-\mu)^{-1}$ where $\mu$ is the chemical potential of the two holes. The whole process of Fig.~\ref{Fig1}b consists of two independent $t_{pd}^2/(\varepsilon_d-\mu)$ hoppings, one per particle, and of the waiting time. Therefore the resulting effective $U_d=\infty$ repulsion $\tilde U$ of two $p$-particles is of the order of

\begin{equation} \tilde U\sim t_{pd}^4/(\varepsilon_d-\mu)^3\label{Eq01}
\end{equation}

\noindent and can be interpreted as a retardation (kinetic) effect.

\section{$U_d=\infty$ slave particle theory of the metallic phase}

We have carried out the corresponding systematic theory which starts from the
$n_d^{(0)}=0$ unperturbed metallic ground state by using the slave particles.
This time dependent diagrammatic approach, of infinite order in the
perturbation $t_{pd}$, requires the use of the spinless fermion-Schwinger
boson representation in order to avoid the degeneracy of the $n_d^{(0)}=0$
unperturbed ground state in the overcomplete slave particle Hilbert space. The
advantage of this representation is that the perturbation theory is manifestly
translationally invariant at each stage, and ultimately locally gauge
invariant. The disadvantage is that the three sorts of particles involved,
$p^\sigma$-fermions, f-spinless fermions and Schwinger´s $b^\sigma$-bosons are
distinguishable, i.e., that the Cu-O anticommutatiom rules are replaced by
commutations. The theory has to be therefore antisymmetrized {\it a
posteriori}. Here, we only quote the results.

First, the omission of the Cu-O anticommutation rules is irrelevant in the
lowest $r=1$ order of the Dyson perturbation theory. The reason is that the Cu
site is initially empty, while one particle on Cu is required for
anticommutation and two for $U_d$ interaction. This makes the $r=1$ expression
for physical single-particle propagators strictly equivalent to the result of
the $t_{pd}$ hybridized HF theory. The HF theory is expressed in terms of two
hybridized propagators, one which starts and finishes with the appropriately
weighted propagation on the O-sites ($pdp$ propagator hereafter) and the other
begins and ends on the Cu-sites ($dpd$ propagator). Both propagators are
characterized by the three $i=L,I,U$ bands of poles (branches) at $\omega_{\vec
k}^{(i)}$ \cite{mr1}. This holds irrespectively of the average HF occupation $n_d^{(1)}$
of the Cu-site associated with the chemical potential $\mu^{(1)}$. On the
other hand, the relations required ultimately by the local gauge invariance
$n_f+n_b=1$ and $n_d=n_b$ are nearly satisfied at $r=1$ only when $n_d^{(1)}$
is small, i.e., the theory then converges quickly. It is therefore important
to keep in mind that \cite{fr2,bs5} $n_d^{(1)}\approx1/2$ at $|x|$ small for
$t_{pd}\gg\Delta_{pd}$ and that finite $t_{pp}$ decreases it \cite{mr1}
further. The overall rule of thumb is that $n_d^{(1)}<1/2$ as long as the HF
chemical potential $\mu^{(1)}$ falls below the vH singularity in the lowest
L-band, i.e., as long as $x<x_{vH}$ where \cite{mr1} $x_{vH}\propto-t_{pp}$ is
the positive doping required to reach the vH singularity. As easily seen,
small $n_d^{(1)}$ corresponds to a weak effective interaction $\tilde
U<t_{pd}$.

The Cu-O (anti)commutation rules are also irrelevant for two further $r =2,3$
expressions. $r =2$ leads to the coherent Brinkmann-Rice-like band narrowing
$t_{pd}^2\rightarrow t_{pd}^2(1+n_d/2)(1-n_d)$ and the variation of
$\Delta_{pd}$ through $\varepsilon_d\rightarrow\tilde\varepsilon_d$. However,
unlike in the mean-field \cite{mr1,ko1,ja1} slave particle approximations, this is accompanied by the incoherent, local, dynamic fluctuations identified as the $d_{10}\leftrightarrow d_9$ charge-transfer disorder. Importantly, the disorder falls far from the Fermi level, although there are indications \cite{ja1,ni1,zl1} that in higher perturbation orders it spreads all over the spectrum.

The interaction $\tilde U$ appears explicitly for $r =4$. It introduces the particle-particle and particle-hole correlations in the single particle $pdp$ and
$dpd$ propagations. $\tilde U>0$ favorizes the coherent particle-hole
correlations which appear here as pseudogaps. The slave fermion theory
predicts however that the effect of $\tilde U$ is the same in the singlet and
the triplet channels, which is the consequence of the omission of the Cu-O
anticommutation rules. The {\it a posteriori} antisymmetrization of the theory
therefore associates $\tilde U$ with the singlet (SDW) scattering only.

The relevant SDW hole-electron correlations are associated to lowest order
with the $pdp-pdp$ bubble, as suggested by Fig.~\ref{Fig1}b, where $\tilde U$
appears as the effective interaction between $p$-particles. In contrast, the
corresponding small $U_d$ theory \cite{fr2,dz2,bs5} involves the $dpd-dpd$
bubble. However, although the spectral densities of the $pdp$ and $dpd$
propagators are complementary, the poles are the same. The associated
elementary intraband bubbles share therefore the properties of the overlap of
the vH singularities and of the (imperfect) Fermi surface (FS) nesting, which
both favor the coherent SDW fluctuations with a dominant $\vec q_{SDW}$.

As mentioned above, $\mu^{(1)}$ in the vicinity of the vH singularity brings
the metallic $U_d=\infty$ theory into the intermediate $\tilde U\approx
t_{pd}$ regime with $n_d^{(1)}\approx1/2$. Various experiments and NQR in
particular \cite{ku1} indicate indeed that for $|x|\leq0.2$  the average
occupation of the Cu site is close to $n_d\approx1/2$. Therefore, we associate
the transition, between the $|x|\approx0$ long- and short-range magnetic order
and the metallic phase at finite $|x|$, to the crossover in $x$ between the
$|x|\approx0$ $t-J$ regime and the finite $|x|$ metallic regime considered
here. This latter is characterized by the close competition of the
$d_{10}\leftrightarrow d_9$  charge-transfer disorder and the coherence
effects which is difficult to cover analytically. In this light, we shall
identify below the physical content of the important hole-electron correlation
functions and determine when their coherent limit is consistent with
experiments in the metallic phase.

\section{Coherent e-h correlations and stripes}

Let us start with the SDW correlations for $x\approx x_{vH}>0$. The vH
overlap/FS nesting behavior of the elementary intraband particle-hole bubble
is not universal. It is however well known that $\GRe\chi_{SDW}^{L,L}(\vec q,
\omega)$  becomes large for $\omega$ small and  close to $\pi/a\;[1,1]$.
Taking formally $t_{pp}=0$, the log square singularity in $\omega$ occurs
\cite{dz2,fr2} at $x=0$ due to the vH overlap and to the perfect 2D FS nesting
at $\vec q_{SDW}= \pi/a\;[1,1]$. Analytical \cite{sc1} and numerical
\cite{tu1} calculations show that imperfect nesting associated with finite $x$
at $t_{pp}=0$ produces spikes at incommensurate $\vec q_{SDW}^{coll}$ along
the zone main axes. With finite $t_{pp}$ the spikes were also obtained
numerically \cite{xu1} for $\vec q^{dg}_{SDW}$ along its diagonals. While it
seems well established \cite{sq1} that the spikes at $\vec q^{coll}_{SDW}$ are
dominated by the pairing of holes and electrons close to the antinodal 
$\pi/a\;[1,0]$ and $\pi/a\;[0,1]$ vH points, the pairing which gives rise to
the spikes at $\vec q^{dg}_{SDW}$ is not yet determined unambiguously.

The single well-defined collinear $\pm\vec q_{SDW}^{coll}$ leg or the nearly circular $|\vec q_{SDW}|$ is observed clearly by the magnetic
neutron scattering \cite{tr2,st1,hi1}. As a rule, $\vec q_{SDW}^{coll}$
appears for small energy transfers while $|\vec q_{SDW}|$ occurs at high
frequencies e.g. in metallic YBCO \cite{st1}. The observation of the strong leading harmonics is consistent with metallic
coherency in the propagation of $t_{pd}$-hybridized particles. There is
however also good NMR evidence \cite{go3}  that non-magnetic disorder is
present in LSCO. It may well correspond, in part, to the local
$d_{10}\leftrightarrow d_9$  charge disorder, which is dynamic in the present
theory but becomes frozen \cite{go3,go1} by the strong coupling \cite{bs2} to
the heavy lattice.

The Emery model also encompasses \cite{bs2} the Cu/O$_2$ and O$_x$/O$_y$
charge transfers and various bond fluctuations within the unit cell. In the
$\vec q\rightarrow0$ limit the D$_4$ symmetry classifies \cite{gz1,ts1,ku3}
those correlations in A$_{1g}$, B$_{1g}$ and B$_{2g}$ irreducible
representations, the bond fluctuations being involved directly \cite{ku3} into
the Raman responses. The elementary bubbles associated with those correlations
differ through coherence factors in their numerators while their respective
poles and integration ranges are the same \cite{ku3}. The $\vec q=0$
B$_{1g,2g}$ modes are dominated \cite{ku3} respectively by the contributions
from the main axes or diagonals of the CuO$_2$ zone. They have their own small
$\vec q $ structures \cite{tu1,xu1} (e.g. the elementary O$_x$/O$_y$ bubble is
logarithmically singular \cite{fr2,bs2} at $x=x_{vH}$). In addition, while the
$\vec q\approx0$ B$_{1g}$ and B$_{2g}$ intracell modes are decoupled from the
$\vec q\approx0$ transfers of the total CuO$_2$ charge among the distant unit
cells \cite{ tu1, bs2,ku3}, the coherent $\vec q\approx0$ Cu/O$_2$ charge
transfer is accompanied \cite{tu1,ku3} by the $\vec q\approx0$ intercell
charge transfer. Only the latter is subject to the long
range Coulomb screening/frustration \cite{tu1,go1,ku3,ki1}.


\begin{figure}[tb]

\begin{center}{\scalebox{0.25}
{\includegraphics{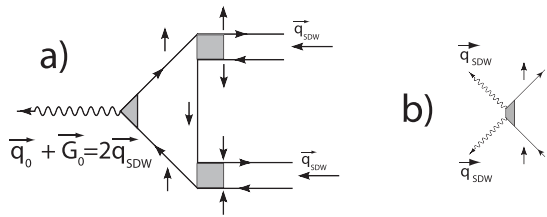}}}
\end{center}

\caption{Umklapp coupling of two SDW's enhanced at $\vec q_{SDW}$ (a) to
intracell charge fluctuations and/or to phonons linearly coupled to carriers;
(b) to phonons coupled quadratically to the carriers. Internal structure of
the electronic triangle is gouverned by the A$_g$, B$_g$ symmetry properties
of the outcoming vertices and by the $U_d$ or $\tilde U$ nature of the
incoming square vertices which flip the spins.\label{Fig2}}

\end{figure}

The small $\vec q$ behavior of the charge transfer correlations is however of
secondary interest if the SDW is taken \cite{sc1} as the dominant fluctuation.
Then, according to Fig.~\ref{Fig2}, two SDW´s enhanced at $\vec
q_{SDW}\approx\pi/a\,[1,1]$ either couple to the intracell fluctuations and
then to phonons, or else directly to phonons. In the case of linear coupling
of Fig.~\ref{Fig2}a the corresponding dominant $\vec q_0$ is small and
satisfies the physically important relation

\begin{equation}
\vec q_0+2\pi/a\;[1,1] =2\vec q_{SDW}\;.\label{Eq02}
\end{equation}

\noindent Alternatively, as in Fig.~\ref{Fig2}b, two such SDW´s can also
couple quadratically \cite{bs2} to tilts of the CuO$_6$ octahedra at $\vec
q_{TILT}=\vec q_{SDW}$. When two SDW´s enhanced at $\vec q_{SDW}^{coll}$ close
to $\pi/a\;[1,1]$ are associated with the $\pi/a\;[1,0], \pi/a\;[0,1]$ pairing
they drive  \cite{ku3}, directly or via the B$_{1g}$ O$_x$/O$_y$ charge transfer, the LTT/(e$_{xx}$-e$_{yy}$) deformations at  $\pm \vec q^{coll}_{TILT}/\pm\vec q_0^{coll}$. Analogously, {\it if} two SDW´s enhanced at $\vec q^{dg}_{SDW}$ close to $\pi/a\;[1,1]$ are associated with the $\pi/a\;[1/2,1/2], \pi/a[-1/2,-1/2]$ pairing they drive \cite{ku3} the B$_{2g}$ O$_x$-O$_y$ bond fluctuations and the LTO/(e$_{xy}$+e$_{yx}$) modes at $\pm \vec
q^{dg}_{TILT}/\pm\vec q_0^{dg}$ .

The $\vec q_{TILT}\approx\pi/a\;[1,1]$ tilts and the $\vec q_0$ modes are in
addition entangled \cite{pu1} by the ionic forces, namely the $\pi/a\;[1,1]$
LTT and LTO tilts are accompanied, respectively, by the {\it homogeneous}
e$_{xx}$-e$_{yy}$ and  e$_{xy}$+e$_{yx}$ shears of the CuO$_2$ planes. Those
entangled single leg deformations thus lift, through the electron-phonon
couplings \cite{bs2}, the degeneracy of two O$_x$/O$_y$ sites \cite{bs2} or of
four $t_{pp}$ bonds within the CuO$_2$ unit cell by dimerizing them two by
two.

The stability of the striped structures can be investigated using the related
\cite{sc1,bs2}  Landau functionals. E.g., the entangled collinear modes are
related in this way to the collinear nematic static stripes
\cite{ki1,em4} usually characterized by $\vec q_0^{coll}$. It appears thus
quite clearly that the coherent O$_x$/O$_y$ charge transfer, coupled to the
lattice, is an essential ingredient of the collinear stripes \cite{ki1} in the
metallic phase. This agrees with observations \cite{tr2,st1,gz1,ts1,bi1,sg1}.
The collinear SDW and O$_x$/O$_y$ charge transfer correlations get enhanced in
metallic lanthanum cuprates for dopings $x\approx x_{vH}$. The spikes
\cite{sq1} at $\vec q^{coll}_{SDW}$ in $\chi^{L,L}_{SDW}(\vec q, \omega)$
then explain the "nematic" version of Eq.~(\ref{Eq02}) observed
\cite{tr2,gz1} in the metallic phase. The corresponding LTO/LTT lattice
instability is predicted \cite{bs2} and observed \cite{ax1,kr1} to be of the
first order in LBCO for $x\approx x_{vH}\approx1/8$. Those effects are
attributed here to the vH overlap/FS nesting, while
the commensurability $1/8$ is expected to play only the secondary role.

In contrast to lanthanum cuprates the Fermi level $\mu$ in the optimally doped
YBCO and BSCO, as measured by ARPES, falls \cite{mr1} well below the vH point,
i.e., $x<x^{HTC}_{vH}$. The collinear SDW-O$_x$/O$_y$ stripes enhanced by the
vH singularities are therefore not expected to occur in metallic YBCO and
BSCO. Indeed, at larger $x$, the collinear stripe structure is replaced \cite{ts1} by the fourfold commensurate stripes in
"checkerboard" configuration \cite{ts1,ha1}, which restores the D$_4$ symmetry.

The salient new feature of the present low-order large $U_d$ analysis is thus
that the coherent, weak, incommensurate collinear \cite{sq1} SDW and
O$_x$/O$_y$ correlations appear to get enhanced with respect to the
$d_{10}\leftrightarrow d_9$ charge disorder if the vH singularities are
reached \cite{in1} by doping $x_{vH}$. This explains the long standing puzzle
\cite{ax1} why the enhanced magnetic/charge coherence occurs twice in LBCO as
a function of doping, once as the long- or short-range \cite{fr3,lo1} AF order
at $x\approx0$ and then again as the incommensurate SDW at single $\pm\vec
q_{SDW}^{coll}$ for finite positive $x_{vH}\approx0.1$, accompanied by the
static O$_x$/O$_y$ charge transfer coupled to the staggered LTT tilting of the
CuO$_4$ octahedra.

\begin{acknowledgments}

Invaluable discussions with J. Friedel, L.P. Gor'kov,
I. Kup\v ci\' c, D.K. Sunko and E. Tuti\v s are gratefully acknowledged. This work was supported by the Croatian Government under Projects No. $119-1191458-0512$ and No. $035-0000000-3187$.

\end{acknowledgments}

\end{document}